\documentclass[prl,twocolumn,groupedaddress]{revtex4}
\usepackage{graphicx,color}
\usepackage{amssymb}   % for math
\usepackage{amsmath}
\usepackage{epstopdf}
\usepackage{natbib}
\usepackage{hyperref}
\usepackage{bm}

\begin{document}
\title{Effects of Domain Walls in Quantum Anomalous Hall Insulator/Superconductor Heterostructures}
\author{Chui-Zhen Chen, James Jun He, Dong-Hui Xu}
\author{K. T. Law} \thanks{phlaw@ust.hk}
\affiliation{Department of Physics, Hong Kong University of Science and Technology, Clear Water Bay, Hong Kong, China. }
\begin{abstract}
In a recent experiment, half-quantized longitudinal conductance plateaus (HQCPs) of height $\frac{e^2}{2h}$ have been observed in quantum anomalous Hall (QAH) insulator/superconductor heterostructure transport measurements. It was predicted that these HQCPs are signatures of chiral Majorana edge states. The HQCPs are supposed to appear in the regimes where the Hall conductance $\sigma_{xy}$ is quantized. However, experimentally, a pair of the HQCPs appear when the Hall conductance $\sigma_{xy}$ is only 80\% of the quantized value when dissipative channels appear in the bulk. The dissipative channels in the bulk are expected to induce Andreev reflections and ruin the HQCPs. In this work, we explain how domain walls can cause $\sigma_{xy}$ to deviate from its quantized value and at the same time maintain the quantization of HQCPs. Our work supports the claim that the experimentally observed HQCPs are indeed caused by chiral Majorana modes in the QAH insulator/superconductor heterostructure. \end{abstract}

\maketitle

{\emph {Introduction.}}--- Majorana fermions are antiparticles of themselves\cite{Wilczek, Kitaev1}. In condensed matter systems, a pair of spatially separated Majorana fermions can appear as a zero energy fermionic excitation in topological superconductors \cite{ Kitaev1}. Importantly, Majorana fermions obey non-Abelian statistics such that braiding Majorana fermions can change the total quantum state of the system \cite{RG, Ivanov, Fujimoto, STF, Alicea2}. As such, Majorana fermions can have potential applications in quantum computations and the search for Majorana fermions has become one of the most important subjects in condensed matter physics \cite{Kitaev2, Nayak,HasanREV,QiREV,FlensbergREV,BeenakkerREV}.

Tremendous progress has been made in the realization and detection of Majorana fermions in the past few years. It was first proposed that the vortex cores of superconducting surface states of topological insulators host Majorana fermions \cite{Fu08}. In a recent experiment, electron tunnelling into the vortex cores using spin polarized scanning tunnelling microscope showed spin dependent conductance \cite{Jia2016} which can possibly be caused by spin filtering effects of Majorana fermions \cite{JamesPRL,BHWu,Oreg,LHH}. Majorana fermions can also appear as end states of superconducting nanowires which possess Rashba type \cite{Alicea2010,Sato2009,ORV,Sau} or Ising type \cite{ZhouTong, Tewari, Aji} spin-orbit coupling. The Majorana fermions can induce zero bias conductance peaks (ZBCP) in tunnelling spectroscopy experiments \cite{Vic2009,Beenakker2} and ZBCPs possibly associated with Majorana fermions have been observed in several experiments \cite{Kouwenhoven,Das,Deng,Yazdani}. However, there is an  on-going debate in the origin of these ZBCPs. %Coulomb blockade effects possibly related to Majorana fermions have also been observed in InAs nanowire experiments [].

\begin{figure}
\centering
\includegraphics[scale=0.55, bb =60 28 650 460, clip=true]{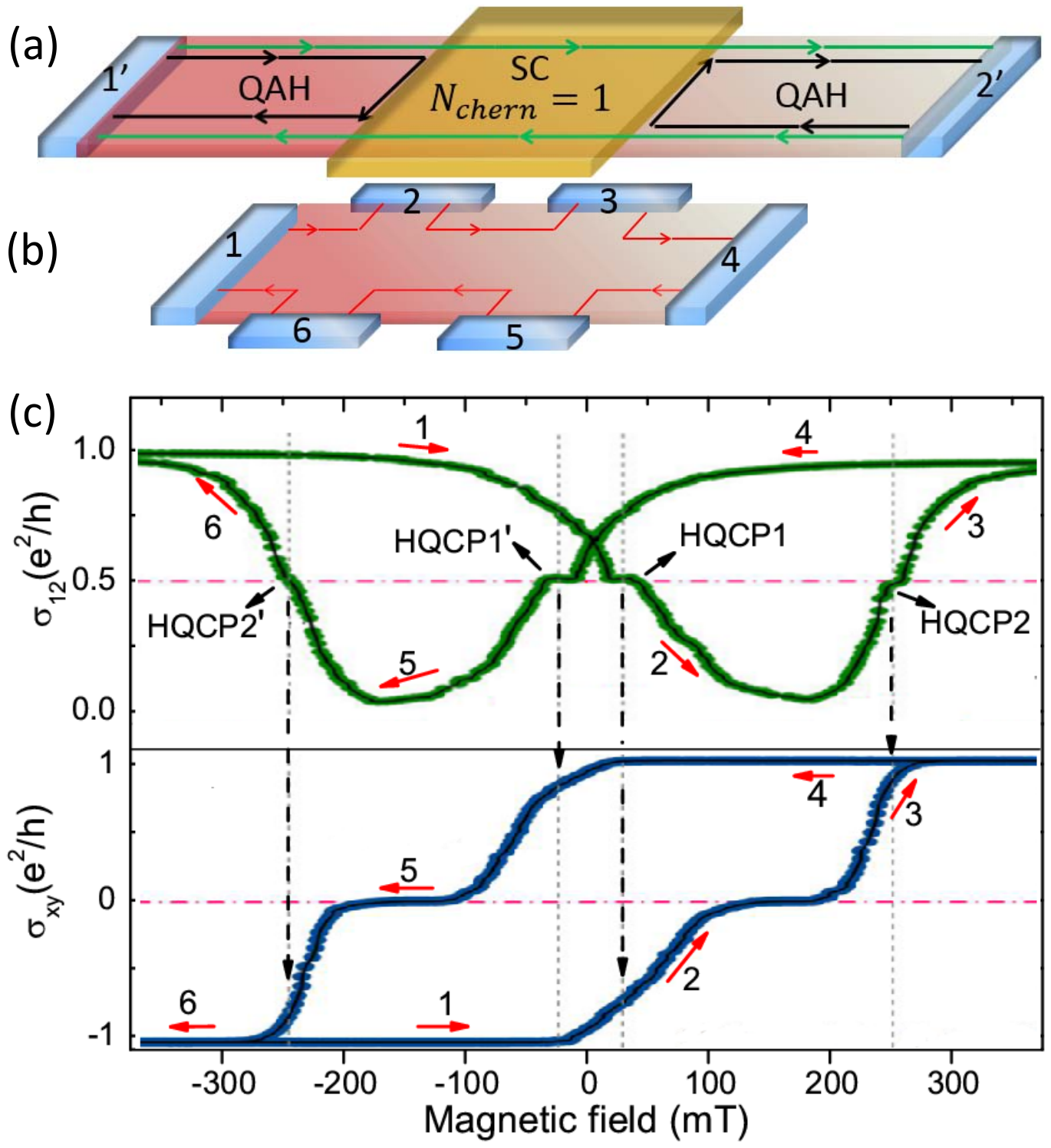}
\caption{(Color online).(a) A QAH insulator/superconductor heterostructure. The QAH insulator under the superconductor becomes a chiral topological superconductor with $N_{Chern}=1$. An incoming fermionic mode from Lead $1'$ is split into two chiral Majorana modes ( green and black lines). Two-terminals conductance is defined as $\sigma_{12}= I/(V_{1'}-V_{2'})$ where $I$ is the current from Lead $1'$ to Lead $2'$ and $V_{i}$ denotes the voltage of Lead $i$. (b) A standard six-terminal Hall bar set-up in the normal state without the superconducting island. (c) Experiment data of $\sigma_{12}$ measured in the superconducting state and $\sigma_{xy}$ measured in the normal state from \cite{Qinglin}. Four $\sigma_{12}$ plateaus with height $\frac{e^2}{2h}$ are observed. They are labeled as HQCP$1$, HQCP$1'$, HQCP$2$ and HQCP$2'$ respectively. HQCP1 and HQCP$1'$ appear when $\sigma_{12} \approx 0.8 e^2/h$. The red arrows indicate the magnetic field sweep directions.
\label{fig1} }
\end{figure}

In 2008, it was proposed by Qi et al. \cite{Qi_QAHSC} that quantum anomalous Hall  (QAH) insulators in proximity to an s-wave superconductor can become a topological superconductor, where a QAH phase can support topologically protected chiral fermionic edge states in the absence of an external magnetic field \cite{Haldane,JingWangREV,LiuREV}. The QAH insulator/superconductor heterostructure can support two topological superconducting phases with Chern number $N_{Chern}$ equals to one or two respectively \cite{Qi_QAHSC,Chung,JamesNC,Sato2,WangJing}. The $N_{Chern}=2 $ superconducting phase is adiabatically connected and topologically equivalent to the QAH insulating phase where a single branch of chiral fermion edge state can be regarded as two branches of chiral Majorana edge states. On the other hand, the $N_{Chern}=1$ phase is a new topological phase which supports one single branch of Majorana fermions propagating at the edges of the sample. It was shown that the two topological superconducting phases with different $N_{Chern}$ have very different transport properties \cite{JamesNC, Sato2}. Interestingly, it was predicted that the two-terminal conductance $\sigma_{12}$ of a QAH insulator, with a superconducting island in the middle (Fig.1a), shows a quantized value at $\frac{e^2}{2h}$, given that the middle island is in the $N_{Chern}=1$ phase which supports chiral Majorana edge modes \cite{Chung,WangJing}.

Experimentally, quantum anomalous Hall phase has been observed \cite{Xue} in magnetic Cr doped Bi$_2$Se$_3$ thin films as predicted previously \cite{Dai}. Quantized Hall resistance $\rho_{xy}$ and almost zero longitudinal resistance $\rho_{xx}$ signifying dissipationless transport has been demonstrated in several recent experiments \cite{Tokura, Kang, Gordon,Kang2,Feng, Chang, Kandala}. However, inducing superconductivity on a QAH insulator to realize the chiral superconducting topological phase is rather difficult as the proximity induced superconducting gap on this magnetic system has to be larger than the bulk gap of the QAH insulator.

Surprisingly, a recent experiment performed by He et al. \cite{Qinglin} observed half-quantized conductance plateaus (HQCPs) with value $\frac{e^2}{2h}$ in a two terminal conductance measurement for experimental set ups depicted in Fig.1a. The two-terminal longitudinal conductance $\sigma_{12}$ data of Ref. \cite{Qinglin} is reproduced in Fig.1c. The normal state (without the superconducting island) Hall conductance $\sigma_{xy}$ data is shown in Fig.1c. The Hall conductance in the normal state is measured in a standard six-probe Hall bar geometry as depicted in Fig.1b.

The appearance of the HQCPs has been predicted previously \cite{WangJing}. Nevertheless, taking a closer look at the experimental data, there are two points worth investigating. First, the HQCPs are supposed to appear when the Hall conductance $\sigma_{xy}$ is quantized \cite{WangJing}. However, from Fig.1c, it is evident that HQCP1 and HQCP$1'$ of $\sigma_{12}$ appear where $\sigma_{xy}$ is only about 80\% of $e^2/h$. Deviation of $\sigma_{xy}$ from the quantized value indicates the presence of dissipative channels in the bulk. Usually, these dissipative channels can introduce extra Andreev reflection channels and ruin the quantization of the HQCPs. So, it is puzzling why the HQCPs are shifted to finite applied magnetic field regime where $\sigma_{xy}$ is not quantized.

Second, HQCP$2$ and HQCP$2'$ in Fig.1c appear when $\sigma_{xy}$ is quantized, and near the trivial to QAH insulator transition points such that QAH bulk gap can be smaller than the induced superconducting gap. The observed locations of HQCP2 and HQCP$2'$ are exactly the same as predicted theoretically \cite{WangJing}. However, HQCP2 and HQCP$2'$ are not flat and change when the external magnetic field changes.

In this work, we focus on explaining why the HQCPs near zero external magnetic field is shifted to the regime where $\sigma_{xy}$ is not quantized. We demonstrate that the shifting of the HQCP1 and HQCP$1'$ are due to the formation of short range domain walls which appear when the external magnetic field changes sign. We found that certain types of domain walls provide extra conducting channels which can give rise to finite value of longitudinal resistance $\rho_{xx}$ but keep the Hall resistance $\rho_{xy}$ quantized. This provides a mechanism to explain the experimental data in Ref.\cite{Qinglin} (Fig.2a). Due to the finite $\rho_{xx}$, $\sigma_{xy}$ deviates from its quantized value even though the chiral edge states are intact. The chiral edge states and the chiral topological superconducting island give rise to HQCPs even when $\sigma_{xy}$ is only about 80\% of its quantized value.

In the following, we first explain how the domain walls can introduce dissipative fermionic modes. Second, we explain how short-range domain walls can give rise to finite longitudinal resistance $\rho_{xx}$ but quantized Hall resistance $\rho_{xy}$ in six-terminal Hall bar measurements. This results in a deviation of $\sigma_{xy}$ from its quantized value. Third, we explain the experimental data by calculating the two-terminal conductance $\sigma_{12}$ in the presence of domain walls.

{\emph{Conducting channels created by domain walls.}}--- Several recent experiments demonstrated that Cr doped Bi$_2$Se$_3$ thin films exhibit quantized Hall resistance $\rho_{xy}$ at zero magnetic field but finite longitudinal resistance $\rho_{xx}$ \cite{Gordon,Kandala,Feng,Kang,Tokura}. The experimental data from Ref.\cite{Gordon} are reproduced in Fig.2a. It is important to note that, when the external magnetic field switches sign, there is a range of magnetic fields in which $\rho_{xx}$ is finite but $\rho_{xy}$ is quantized. As explained in the next section, this is caused by the presence of domain walls which introduce dissipative conducting channels in the system without affecting the quantization of $\rho_{xy}$. To understand the effect of domain walls, we first study the effective Hamiltonian describing a Cr doped Bi$_2$Se$_3$ thin film \cite{WangJing3}.

\begin{figure}
\centering
\includegraphics[scale=0.35, bb = 40 30 800 380, clip=true]{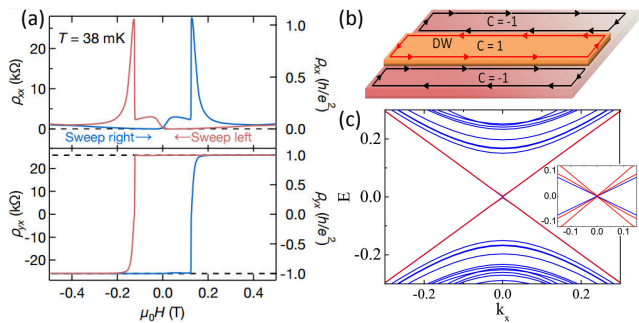}
\caption{(Color online). (a) Experimental data of $\rho_{xx}$ and $\rho_{xy}$ of Cr doped Bi$_2$Se$_3$ taken from \cite{Gordon}. Near the quantum phase transition ($H\sim0$), $\rho_{xy}$ is quantized while $\rho_{xx}$ is nonzero. (b) When a domain wall with $N_{Chern}=-1$ is formed in QAH insulator with $N_{Chern}=1$, the double chiral fermionic edge modes emerge at the boundary. (c) Energy dispersion of the QAH with domain wall along $x$-direction and double chiral edge modes are labeled as red lines. The insert is the zoom-in image of the chiral edge states near $k_x=0$, the Fermi velocity of the domain wall is artificially changed to make the double chiral edge modes more visible.
\label{fig2} }
\end{figure}
The effective Hamiltonian of the QAH insulator in real space can be written as:
\begin{eqnarray}
% \nonumber to remove numbering (before each equation)
  H_{eff} &=&
    \int d\textbf{r} \Psi^{\dagger}(\textbf{r})\{v_F \hat{k}_y\tau_z\otimes \sigma_x
     - v_F \hat{k}_x \tau_z \otimes \sigma_y \nonumber\\
  & &+ m(\hat{\textbf{k}}) \tau_y + M'_z \tau_z\otimes\sigma_z  \} \Psi(\textbf{r}).
  \label{H_e}
\end{eqnarray}

Here, $\Psi(\textbf{r}) =[\psi_{t\uparrow},\psi_{t,\downarrow},\psi_{d\uparrow},\psi_{d,\downarrow}]^{T} $
is a four component electron operator, where $t$ and $b$ label electrons from the top and bottom layers of the topological insulator surfaces respectively. $\uparrow$ and $\downarrow$ are the spin indices. The Pauli matrices $\sigma_{x,y,z}$ and $\tau_{y,z}$ are defined in spin and layer space, respectively.
$v_F$ is the Fermi velocity and $m(\hat{\textbf{k}})= m_0 - 2m_1(\hat{k}^2_x + \hat{k}^2_y)$ describes the coupling between the top  layer and bottom layer, while $M'_z$ is the magnetization in z direction induced by the magnetic impurities \cite{WangJing3}. For simplicity, we set $v_F=1$ and $m_1=1$.
When $ M'_z> |m_0|$, the system is in the QAH phase with $N_{Chern}=1$. When $M'_z $ changes sign, the Chern number changes sign as well. It is reasonable to expect domain walls to be formed when the external magnetic field changes sign and weaker than the coercive field.

To study how conducting channels can be created by forming domain walls, we consider a strip of the QAH system parallel to the $x$-direction which has opposite magnetization direction compared to the rest of the bulk as depicted in Fig.2b. Further assuming periodic boundary conditions in the $x$-direction, the energy spectrum of the whole system is shown in Fig.2c. It is evident from Fig.2c (and schematically shown in Fig.2b) that the middle domain wall creates double chiral fermionic edge modes at the domain wall boundary. Double chiral fermionic modes are expected because the domain wall has opposite Chern number than the bulk. Therefore, quite generally, domain walls introduce conducting channels in the system. These conducting channels are different from the quasi-helical modes discussed in Ref.\cite{WangJing4} which are inherited from the band structure of the topological insulator thin film.  Those quasi-helical modes are not affected by magnetic fields.

{\emph{Quantized $\rho_{xy}$ and finite $\rho_{xx}$}}--- A standard six-terminal Hall bar geometry is depicted in Fig.1b. The Hall conductance is defined as $\rho_{xy} = (V_{2}-V_{6})/I$ where $I$ is the current that goes from the source with voltage $V_1$ to the drain with voltage $V_4$. $V_{i}$ denotes the voltage of lead $i$ defined in Fig.1c. According to the Landauer-Buttiker formula \cite{Buttiker}, $I_{i} = \frac{e^2}{h} (\sum_{k} T_{ik} V_{k}- T_{ki}V_i)$ where $I_{i}$ is the current flowing out of lead $i$ and $T_{ik}$ is the transmission coefficient from lead $k$ to lead $i$. Using the tight-binding version of $H_{eff}$ in Eq.{\ref{H_e}} \cite{supp}, $T_{ij}$ can be calculated using the recursive Green's function method \cite{PatrickGF}, where
\begin{equation}
T_{ij} = \text{Tr}[\Gamma_{i}G^{r}({\bf r} )\Gamma_{j}G^a({\bf r})].
\end{equation}
Here, $G^{r,a}(E)=[E-H_{eff}-\sum_{i=1}^{6}\Sigma^{r,a}_{i}]^{-1}$ are retarded (advanced) Green function  and $\Gamma_{i}=i(\Sigma^{r}_{i} -\Sigma^{a}_{i})$ with $\Sigma^{r,a}_{i}$ the self-energy of the $i$th lead. Using the recursive Green's function method to calculate $T_{ij}$ allows us to calculate $\rho_{xy}$ and $\rho_{xx}$ even in the presence of complicated domain walls. In the uniform phase without domain walls, and when the system is in the QAH phase with $N_{Chern}=1$, by fixing $I_{1}=-I_{4}$ and solving the Landauer-Buttiker formula, we have $V_2=V_3=V_1$ and $V_4=V_5=V_6$. This gives $\rho_{xy}= (V_{2}-V_{6})/I = h/e^{2}$ and $\rho_{xx}=0$ as expected \cite{WangJing}. Equivalently, we have $\sigma_{xy}=\rho_{xy}/(\rho_{xy}^2 + \rho_{xx}^2) = e^{2}/h$ and $\sigma_{xx}=\rho_{xx}/(\rho_{xy}^2 + \rho_{xx}^2) = 0$ as shown in Fig.3e.

Interestingly, in Cr doped $•$ thin films, it was shown experimentally that there are regimes in which $\rho_{xy}$ is quantized at $h/e^{2}$ but $\rho_{xx}$ is finite as shown in Fig.2a \cite{Gordon}. This happens when the external magnetic field changes sign but is smaller than the coercive field. We expect domain walls to be created when the external magnetic field starts to flip the magnetization direction of the sample.

To be specific, we add a domain wall connecting Lead 3 and Lead 4 as shown in Fig.3a. As explained in the last section, we expect the  domain wall to introduce double chiral channels going from Lead 3 to Lead 4 as shown in Fig.3a. Importantly, this domain wall also introduces a single chiral channel (the edge mode between the domain wall and the vacuum) which goes from Lead 4 to Lead 3. Due to the double chiral channel connecting Lead 3 to Lead 4, $T_{34}$ is changed from $1$ to $2$ in the Landau-Buttiker formula. On the other hand, $T_{43}$ is changed from $0$ to $1$ because of a chiral fermionic mode shown in Fig.3a. These changes of transmission coefficients $T_{ij}$ from physical arguments can easily be checked by numerical calculations as shown in Fig.3c. 

To simulate the effects of the coercive field, we assume the magnetization of the bulk QAH insulator to be $M'_z= M_z-0.18$ in Figs.3c to 3e, where $M_z$ is linear proportional to the external magnetic field. Therefore, without domain walls, $T_{43}=1$ and $T_{34}=0$ when $M_z \approx 0.1$ in Fig.3c. This indicates that electrons propagate from Lead 3 to Lead 4 through a chiral fermionic mode (such that the current goes from Lead 4 to Lead 3). The magnetization of the domain wall in Figs.3c to 3e is set to be $M_z + 0.12$. For large negative $M_z$, both the domain wall and the bulk have $N_{Chern}=-1$ and the domain wall has no significant effects. Near $M_z \approx 0$, the domain wall has opposite magnetization as the bulk. In this case, $T_{43}$ can reach about 2 ( due to the double chiral fermionic modes as shown Fig.3a). At the same time, $T_{34}$ can increase from 0 to 1 (due to the chiral mode  between the domain wall and vacuum as shown in Fig.3a). Near $M_z=0.2$, the bulk is topologically trivial but the domain wall has $N_{Chern}=1$. Therefore, $T_{43}=T_{34}=1$. For large and positive $M_z$, the domain wall and the bulk has $N_{Chern}=1$. As a result, $T_{43}=0$ and $T_{34}=1$ as the chiral fermionic mode changes its propagation direction compared to the $N_{Chern}=-1$ case with large negative $M_z$.

Due to the change in $T_{ij}$ near $M_z \approx 0$, the voltage in Lead 3 is lower than Lead 2. This results in finite $\rho_{xx}$ near $M_z \approx 0$ as shown in Fig.3d. It is important to note that $\rho_{xy}$ is still quantized at $h/e^2$ even though $\rho_{xx}$ is finite. This is one of the main results of this work which can explain the experimental observation in Ref.\cite{Gordon}.

Importantly, the details, such as the shape of the domain wall, are not important as long as conducting channels can be introduced connecting Lead 3 and Lead 4. Other domain walls, which are isolated in the bulk or connecting leads with the same voltages (such as Lead 5 and Lead 6), do not affect the transport of the system as discussed in the Supplementary Material \cite{supp}. Therefore, the situation studied in this section is quite general. 

\begin{figure}
\centering
\includegraphics[scale=0.42, bb = 50 30 800 580, clip=true]{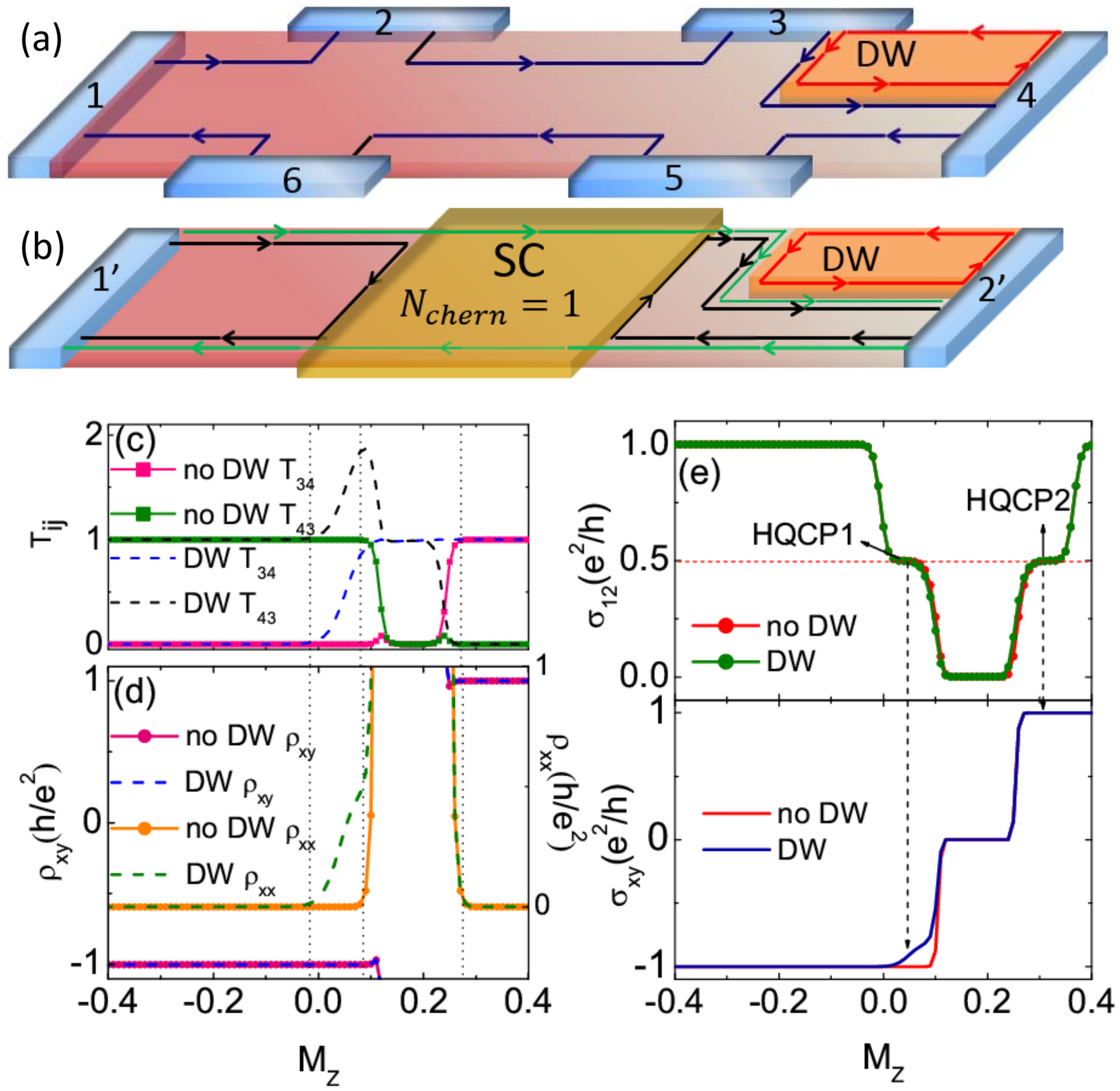}
\caption{(Color online).(a) Schematic plot of the six-terminal Hall-bar measurement of QAH with a domain wall (DW).
%The sizes of Hall-bar are $W=100$, $L_1=200$, $L_2=50$, where QAH/SC sample has the same size with $L_s=300$. The size of domain wall is fixed to be %$20\times60$.
 (b)  The two-terminal measurement of QAH insulator/superconductor heterostructure with a DW.  (c) $T_{34}$ and $T_{43}$ dependence on $M_z$ in setup (a). (d) $\rho_{xx}$ and $\rho_{xy}$  in the presence and absence of DWs. With DWs, $\rho_{xx}$ is finite when $\rho_{xy}$ is quantized as observed in the experiment near $M_z \approx 0$. (e) The two-terminal conductance $\sigma_{12}$ and Hall conductance $\sigma_{xy}$ as a function of $M_z$. $\sigma_{12}$ is not affected by the DWs. Without DWs, HQCP1 appear only when $\sigma_{xy}$ is quantized. In the presence of DWs, the HQCP1 appears when $\sigma_{xy}$ is not quantized as observed in the experiment. On the other hand, HQCP2 appear when $\sigma_{xy}$ is quantized.
\label{fig3} }
\end{figure}

{\emph{Half-quantized $\sigma_{12}$}.}--- In proximity to an s-wave superconductor, QAH becomes a topological superconductor and the Bogoliubov-de Genne Hamiltonian can be written as
\begin{eqnarray}
% \nonumber to remove numbering (before each equation)
  H_{BdG} &=& \left(
                \begin{array}{cc}
                  H_{eff}(k)-\mu & \Delta \\
                  \Delta^{\dagger} & -H_{eff}^{\ast}(-k)+\mu \\
                \end{array}
              \right)
   \\
  \Delta &=& \left(
               \begin{array}{cc}
                 \Delta_{t}i\sigma_y & 0 \\
                 0 & \Delta_{b}i\sigma_y \\
               \end{array}
             \right),  \nonumber
\end{eqnarray}
where the $\Delta_{t}$ and $\Delta_{b}$ are induced pairing potential of top and bottom layers, respectively and $\mu$ is the chemical potential.
As explained in Refs. \cite{Qi_QAHSC, Chung, WangJing, JamesNC}, when the condition $\Delta_{t}=-\Delta_{b} > |m_{0}-M'_z|$ is satisfied, the QAH state is turned into a chiral topological superconductor with $N_{Chern}=1$ and a single branch of chiral Majorana modes. Since the QAH insulator is topologically equivalent to the chiral topological superconducting phase with $N_{Chern}=2$, there must be a chiral Majorana mode at the boundary between the QAH phase and the chiral topological superconducting phase with $N_{Chern}=1$ as depicted in Fig.1a. As a result, when a chiral fermionic mode of the QAH phase is injecting electrons into the $N_{Chern}=1$ chiral topological superconductor, the chiral fermionic mode is split into two branches of Majorana modes as shown in Fig.1a \cite{Chung, WangJing}. One branch of the Majorana mode propagates through the $N_{Chern}=1$ chiral topological superconductor. Since Majorana modes are linear combinations of electrons and holes, this results in equal normal tunnelling and crossed Andreev reflection probabilities. The other branch of Majorana propagates along the interface between the QAH insulator and the chiral topological superconductor. This results in equal normal reflection and Andreev reflection probabilities. Therefore, the probability of all these processes are equal to 1/4 \cite{Chung, WangJing}. When this happens, the two-terminal conductance $\sigma_{12} = (V_1'-V_2')/I $ has a quantized value of $\frac{e^2}{2h}$ \cite{Chung, WangJing}. The experimental set up is depicted in Fig.1a and similarly in Fig.3b. The fixed probability of 1/4 for all the four tunnelling processes depends on the assumption that there are no other conducting channels in the bulk.

However, experimentally, the HQCPs appear when the Hall conductance $\sigma_{xy}$ is only about 80\% of the quantized value and such a large shift to the non-quantized regime cannot be caused by finite temperature effects.. This is a rather surprising result. Non-quantized value of $\sigma_{xy}$ indicates the appearance of bulk states which can introduce extra transport channels and these channels can ruin the quantization of the HQCP.

To understand the experimental results, we note that $\sigma_{xy}=\rho_{xy}/(\rho_{xy}^2 + \rho_{xx}^2)$. From Fig.3e, $\sigma_{xy}$ is not quantized if $\rho_{xx}$ is finite, even when $\rho_{xy}$ is quantized from Fig.3d. In the presence of domain walls as depicted in Fig.3a, the chiral edge states in most part of the sample are not affected by the domain wall. As a result, the two-terminal conductance $\sigma_{12}$ can still be quantized. To understand the quantization of $\sigma_{12}$, we introduce pairing terms in the middle region of the QAH system as depicted in Fig.3b. The two terminal conductance $\sigma_{12}$ is calculated as a function of $M_z$ which is linearly proportional to the applied magnetic field  both in the presence and in the absence of domain walls. The results are depicted in Fig.3e. It is clear that, in the absence of domain walls, the HQCPs have to appear when $\sigma_{xy}$ is quantized. However, in the presence of domain walls, HQCP1 appears when $\sigma_{xy}$ is not quantized as observed in the recent experiment \cite{Qinglin}.

On the other hand, HQCP2 appears when $\sigma_{xy}$ is quantized and near the trivial insulator to QAH insulator transition point. This is because, at field strengths which are higher than the coersive field of about 150mT \cite{Qinglin}, the whole system enter a single domain regime. Unfortunately, HQCP2 is not flat and its value depends on the magnetic field strength. This is another question which deserves further investigation.

{\emph{Conclusion}} --- In this work, we show that domain walls which induce dissipative channels can give rise to finite $\rho_{xx}$ but quantized $\rho_{xy}$. These domain walls allow the HQCPs to appear in the regime where $\sigma_{xy}$ is non-quantized. Our work supports the claim that the experimentally found HQCPs are indeed due to chiral Majorana modes of the $N_{Chern}=1$ chiral topological superconductor as predicted in Ref.\cite{WangJing}.

%On the other hand, HQCP2 and HQCP2' appear when $\sigma_{xy}$ is quantized but they also has linear dependence on the external field. This dependence can be caused by the orbital effects of the external magnetic field in high field regime when the magnetic field is stronger than the $H_{c1}$ of $Nb$. The orbital effects are ignored in previous and our current analysis which give rise to quantized HQCP2 and HQCP2'.

{\emph{Acknowledgement.}}--- KTL thanks Qing-Lin He, Lei Pan and Kang Wang for illuminating discussions and hosting him at UCLA where part of this work was done. The authors thank the support of HKRGC and Croucher Foundation through HKUST3/CRF/13G, 602813, 605512, 16303014 and Croucher Innovation Grant.

\end{document}